\documentclass[twocolumn,prl,aps,showpacs,groupedaddress]{revtex4}

\usepackage{epsfig}

\begin{document}

\title{Anomalous Diffusion of Inertial, Weakly Damped Particles}

\author{R.~Friedrich, F.~Jenko$^1$, A.~Baule$^2$, S.~Eule}
\affiliation{Institute for Theoretical Physics, University of M\"unster,
Wilhelm-Klemm-Str. 9, D-48149 M\"unster, Germany\\
$^1$Max-Planck-Institut f\"ur Plasmaphysik,
Boltzmannstr. 2, D-85748 Garching, Germany\\
$^2$Department of Physics and Astronomy,
University of Leeds, Leeds LS2 9JT, UK} 
\date{\today}



\begin{abstract}

The anomalous (i.e. non-Gaussian) dynamics of particles subject to a deterministic
acceleration and a series of 'random kicks' is studied. Based on an extension of the
concept of continuous time random walks to position-velocity space, a new fractional
equation of the Kramers-Fokker-Planck type is derived. The associated collision
operator necessarily involves a fractional substantial derivative, representing
important nonlocal couplings in time and space. For the force-free case, a closed
solution is found and discussed.

\end{abstract}

\maketitle

For more than 100 years, the quest to understand the dynamics of tracer particles
in random environments has both challenged and inspired countless scientists. Many
fundamental questions about the dynamical aspects of statistical physics have been
raised in this context, and most of them are still unanswered today. Given the fact
that this nonequilibrium phenomenon plays a prominent role in such diverse areas as
condensed matter physics, biophysics, physical chemistry, astrophysics, plasma physics,
and turbulence research, it is not at all surprising that its theoretical description
has continued to attract a lot of attention over all these years.

In the more recent past, a strong emphasis in this area of research has been on the
investigation of the origin and the consequences of anomalous, i.e. non-Gaussian,
diffusion (see, e.g., \cite{Shlesinger,Zaslavsky,Metzler}).
While most publications on this topic are based on a random-walk type
description in real space, this Letter employs a phase (position-velocity) space
approach, rigorously deriving and analyzing a new fractional equation of the
Kramers-Fokker-Planck (KFP) type. For later reference, we mention here that the
usual KFP equation is given by \cite{Risken,vanKampen}
\begin{equation}\label{Kramers}
   \left[{\partial\over\partial t}+{\bf u}\cdot\nabla_x
   +{\bf A}({\bf x})\cdot\nabla_u\right]
   f({\bf x},{\bf u},t) = {\cal L}_{\rm FP}f({\bf x},{\bf u},t)
\end{equation}
where we have introduced the Fokker-Planck collision operator
\begin{equation}\label{FPop}
   {\cal L}_{\rm FP}f=\gamma\nabla_u\cdot({\bf u}f)+\kappa\,\Delta_u f\,.
\end{equation}
As Kramers \cite{Kramers} has shown, Eq.~(\ref{Kramers}) determines the time evolution
of the joint position-velocity probability distribution $f({\bf x},{\bf u},t)$ of a
damped particle with inertia in a force field under the influence of an additional
fluctuating force with the usual white noise statistics:
\begin{equation}\label{dyn1}
  \dot{\bf x}(t) = {\bf u}(t)\,,\quad
  \dot{\bf u}(t) = {\bf A}({\bf x})-\gamma{\bf u}(t)+\mbox{\boldmath$\Gamma$(t)}
\end{equation}
with
\begin{equation}\label{dyn2}
  \langle\Gamma_i(t)\Gamma_j(t')\rangle = 2\kappa\,\delta_{ij}\delta(t-t')\,.
\end{equation}
In 1965, Montroll and Weiss \cite{Montroll} introduced a different class of stochastic processes
denoted as continuous time random walks (CTRWs). -- For a recent, comprehensive review, the reader
is referred to Ref.~\cite{Metzler}. -- Conventional CTRWs are linked to Eq.~(\ref{dyn1})
in the limit of large damping $\gamma$ and vanishing acceleration ${\bf A}({\bf x})$. Here, the
dynamics is described by the Langevin equation $\dot{\bf x}(t)=\mbox{\boldmath$\Gamma$}(t)/\gamma$.
In practice, however, many systems in physics, chemistry, and biology will not satisfy this condition.
Thus we are led to consider the CTRW analog of Eq.~(\ref{dyn1}).

The dynamical model we propose contains both deterministic and stochastic
elements. The motion of an individual particle is assumed to be governed
by the time evolution equations
\begin{equation}\label{dyn}
   \dot{\bf x}(t)={\bf u}(t)\,,\quad \dot{\bf u}(t)={\bf A}({\bf x})\,.
\end{equation}
However, from time to time, the particle is subject to a 'random kick' which
changes its velocity abruptly. This kind of dynamics can be described by means
of a CTRW-like approach in phase space.
Let us consider a particle which at time $t'$ is located in the volume
element $d{\bf x}'$ about ${\bf x}'$, and in the time interval $[t',t'+dt']$
changes its velocity to a new value which lies in the velocity space element
$d{\bf u}'$ about ${\bf u}'$. The probablity for such a process shall be
denoted by
\begin{equation}
   \eta({\bf x}',{\bf u}',t')\,d{\bf x}'\,d{\bf u}'\,dt'\,.
\end{equation}
After a (random) time period $\tau=t-t'$, this particle will undergo a further 
transition to a state with the velocity ${\bf u}$ at the position ${\bf x}$.
The corresponding conditional probability shall be called
\begin{equation}
   \xi({\bf x},{\bf u},\tau;{\bf x}',{\bf u}')\,d{\bf x}\,d{\bf u}\,d\tau\,.
\end{equation}
We assume that this quantity can be written in the form
\begin{eqnarray}\label{prob}
   & & \xi({\bf x},{\bf u},\tau;{\bf x}',{\bf u}') = W(\tau)\,
   \delta({\bf x}-{\bf X}_\tau({\bf x}',{\bf u}')) \nonumber \\
   & & \times\int d{\bf u}''\,F({\bf u};{\bf u}'')\,
   \delta({\bf u}''-{\bf U}_\tau({\bf x}',{\bf u}'))\,.
\end{eqnarray} 
Here, $W(\tau)\,d\tau$ describes the probability that a transition occurs in
the time interval $[\tau,\tau+d\tau]$, and $F({\bf u};{\bf u}'')\,d{\bf u}$
gives the probability that the particle's velocity will end up in the velocity
space element $d{\bf u}$ about ${\bf u}$. Moreover,
\begin{equation}\label{mapping}
   {\bf x} = {\bf X}_\tau({\bf x}',{\bf u}')\,,\quad
   {\bf u} = {\bf U}_\tau({\bf x}',{\bf u}')
\end{equation}
is the solution (at time $t$) of Eq.~(\ref{dyn}) with initial values (at time $t'$)
${\bf x}'$ and ${\bf u}'$. Assuming that the
distribution function $\xi({\bf x},{\bf u},\tau;{\bf x}',{\bf u}')$ is
statistically independent from the particle's path and using Eq.~(\ref{prob}),
we can relate the quantities $\eta({\bf x},{\bf u},t)$ and
$\eta({\bf x}',{\bf u}',t')$ via the equation
\begin{eqnarray}\label{eq7}
   & & \eta({\bf x},{\bf u},t) - f_0({\bf x},{\bf u})\,\delta(t) = \nonumber \\
   & & = \int_0^t dt' \int d{\bf u}' \int d{\bf x}' \,
   \xi({\bf x},{\bf u},t-t';{\bf x}',{\bf u}')\,
   \eta({\bf x}',{\bf u}',t') = \nonumber \\
   & & = \int_0^t dt'\,W(t-t')\int d{\bf u}'\,
   F({\bf u};{\bf u}')\,{\cal P}^{t,t'}\eta({\bf x},{\bf u}',t')\,.
\end{eqnarray}
Here, $f_0({\bf x},{\bf u})$ characterizes the initial condition, and the
deterministic evolution is described by the Perron-Frobenius operator
\begin{eqnarray}
  & & {\cal P}^{t,t'}\eta({\bf x},{\bf u},t') = e^{-(t-t')[{\bf u}\cdot\nabla_x+
  {\bf A}({\bf x})\cdot\nabla_u]}\,\eta({\bf x},{\bf u},t') = \nonumber \\
  & & = \int d{\bf u}' \int d{\bf x}' \,
  \delta({\bf x}-{\bf X}_\tau({\bf x}',{\bf u}'))
  \delta({\bf u}-{\bf U}_\tau({\bf x}',{\bf u}')) \nonumber \\
  & & \times\,\eta({\bf x}',{\bf u}',t')\,.
\end{eqnarray}
Note that, alternatively, one can write
\begin{equation}
  {\cal P}^{t,t'}\eta({\bf x},{\bf u},t') = {\eta({\bf X}_\tau^{-1}({\bf x},{\bf u}),
  {\bf U}_\tau^{-1}({\bf x},{\bf u}),t')\over |J_\tau({\bf x},{\bf u})|}
\end{equation}
where $({\bf X}_\tau^{-1},{\bf U}_\tau^{-1})$ denotes the inverse of the mapping
(\ref{mapping}), and $J_\tau$ is the corresponding Jacobian. [For a Hamiltonian
dynamical system, $J_\tau$ equals unity.]

Having established an integral equation which determines the time evolution of
$\eta({\bf x},{\bf u},t)$, we are now interested in the joint position-velocity
distribution function $f({\bf x},{\bf u},t)$ which is defined as
\begin{equation}\label{eq8}
   f({\bf x},{\bf u},t) = \int_0^t dt' \, w(t-t') \,
   {\cal P}^{t,t'}\eta({\bf x},{\bf u},t')\,.
\end{equation}
Here, $w(\tau)$ denotes the probability that no random kick occurs within the
time interval $\tau$. For this quantity we have the obvious relationship
\begin{equation}\label{eq9}
   w(\tau) = 1 - \int_0^\tau dt \, W(t)\,.
\end{equation}
In order to derive an equation describing the time evolution of
$f({\bf x},{\bf u},t)$, we first write down the Laplace transforms of
Eqs.~(\ref{eq7}) and (\ref{eq8}),
\begin{eqnarray}\label{eq10}
   & & \eta({\bf x},{\bf u},\lambda) = f_0({\bf x},{\bf u})
   + \int d{\bf u}' \, F({\bf u};{\bf u}') \nonumber \\
   & & \times\,W(\lambda+{\bf u}'\cdot \nabla_x+
   {\bf A}({\bf x})\cdot\nabla_{u'})\,\eta({\bf x},{\bf u}',\lambda)
\end{eqnarray}
and
\begin{equation}\label{eq11}
   f({\bf x},{\bf u},\lambda) = w(\lambda+{\bf u}\cdot\nabla_x+
   {\bf A}({\bf x})\cdot\nabla_u)\,\eta({\bf x},{\bf u},\lambda)\,.
\end{equation}
Since in Laplace space, Eq.~(\ref{eq9}) reads
\begin{equation}\label{eq12}
   w(\lambda) = {1-W(\lambda)\over\lambda}\,,
\end{equation}
Eq.~(\ref{eq11}) can be rewritten as
\begin{eqnarray}\label{eq13}
   & & (\lambda+{\bf u}\cdot\nabla_x+{\bf A}({\bf x})\cdot\nabla_x)\,
   f({\bf x},{\bf u},\lambda) = \nonumber \\
   & & = [1-W(\lambda+{\bf u}\cdot\nabla_x+{\bf A}({\bf x})\cdot\nabla_u)]\, 
   \eta({\bf x},{\bf u},\lambda)\,.\quad
\end{eqnarray}
Eqs.~(\ref{eq10}) and (\ref{eq13}) then yield
\begin{eqnarray}\label{zus}
   & & (\lambda + {\bf u}\cdot\nabla_x + {\bf A}({\bf x})\cdot\nabla_u)\,
   f({\bf x},{\bf u},\lambda) = \nonumber \\
   & & = f_0({\bf x},{\bf u}) + \int d{\bf u}'\,F({\bf u};{\bf u}')\,
   \Phi(\lambda+{\bf u}'\cdot\nabla_x+{\bf A}({\bf x})\cdot\nabla_{u'})
   \nonumber \\
   & & f({\bf x},{\bf u}',\lambda) -
   W(\lambda+{\bf u}\cdot\nabla_x+{\bf A}({\bf x})\cdot\nabla_u)\,
   \eta({\bf x},{\bf u},\lambda)
\end{eqnarray}
where we have introduced the quantity
\begin{equation}\label{eq14}
   \Phi(\lambda)={\lambda\,W(\lambda)\over 1-W(\lambda)}=
   {1-\lambda\,w(\lambda)\over w(\lambda)}\,.
\end{equation}
Expressing $f_0$ in terms of $f$ as
\begin{equation}
   f_0({\bf x},{\bf u}) = f({\bf x},{\bf u},t=0)
\end{equation}
and using the identity
\begin{eqnarray}
   & & W(\lambda+{\bf u}\cdot \nabla_x+{\bf A}({\bf x})\cdot\nabla_u)\,
   \eta({\bf x},{\bf u},\lambda) = \nonumber \\
   & & = \Phi(\lambda+{\bf u}\cdot \nabla_x+{\bf A}({\bf x})\cdot\nabla_u)\,
   f({\bf x},{\bf u},\lambda)
\end{eqnarray}
which follows from Eqs.~(\ref{eq13}) and (\ref{eq14}), the Laplace inversion
of Eq.~(\ref{zus}) yields
\begin{eqnarray}\label{master}
   & & \left[{\partial\over\partial t}+{\bf u}\cdot\nabla_x
   +{\bf A}({\bf x})\cdot\nabla_u\right]
   f({\bf x},{\bf u},t) = \nonumber \\
   & & = \int_0^t dt'\,\Phi(t-t')\,\int d{\bf u}'\,F({\bf u};{\bf u}')\,
   {\cal P}^{t,t'} f({\bf x},{\bf u}',t') - \nonumber \\
   & & -\int_0^t dt'\,\Phi(t-t')\,{\cal P}^{t,t'} f({\bf x},{\bf u},t')\,.
\end{eqnarray}
Thus the time evolution of $f({\bf x},{\bf u},t)$ is given by this master equation.
In the language of kinetic theory, Eq.~(\ref{master}) can be interpreted as follows.
The left-hand side describes a system of particles subject to Eq.~(\ref{dyn}). The
right-hand side represents a collision operator which consists of a source and a sink.
The phase space density of particles at $({\bf x},{\bf u})$ is increased at time $t$
by particles starting from ${\bf X}_\tau^{-1}({\bf x},{\bf u})$ at time $t'$ with
a velocity ${\bf U}_\tau^{-1}({\bf x},{\bf u})$ and making a transition to the
velocity ${\bf u}$ at time $t$ and position ${\bf x}$. $f({\bf x},{\bf u},t)$ is
decreased, on the other hand, by particles making a transition from the velocity
${\bf u}$ to some other velocity. In this context, it should be pointed out that
\begin{eqnarray}
   & & \int_0^t dt'\,\Phi(t-t')\,\int d{\bf u}''\,F({\bf u}'';{\bf u})\,
   {\cal P}^{t,t'} f({\bf x},{\bf u},t') = \nonumber \\
   & & = \int_0^t dt'\,\Phi(t-t')\,{\cal P}^{t,t'} f({\bf x},{\bf u},t')
\end{eqnarray}
due to the constraint
\begin{equation}\label{constr}
   \int d{\bf u}''\,F({\bf u}'';{\bf u}) = 1\,,
\end{equation}
and that the Laplace inversion of Eq.~(\ref{eq14}) yields
\begin{equation}\label{kernel}
   {dw(t)\over dt} = -\int_0^t dt'\,\Phi(t-t')\,w(t')
\end{equation}
which identifies $\Phi(t)$ as a kernel determining $w(t)$. Also note that
${\cal P}^{t,t'} f({\bf x},{\bf u},t')$ is simply a solution of the collisionless
equation
\begin{equation}
   \left[{\partial\over\partial t}+{\bf u}\cdot\nabla_x
   +{\bf A}({\bf x})\cdot\nabla_u\right]f({\bf x},{\bf u},t) = 0\,.
\end{equation}
Obviously, the above collision
operator is highly nonlocal in space and time, in stark contrast to virtually
all expressions commonly used in the kinetic theory of gases or plasmas. This
nonlocality can also be viewed as a retardation effect and is closely linked
to the fact that Eq.~(\ref{master}) is invariant with respect to Galilean
transformations, provided $F({\bf u};{\bf u}')$ depends only on the velocity
difference ${\bf u}-{\bf u}'$. [Galilean invariance implies that together with
$f({\bf x},{\bf u},t)$, every $f({\bf x}-{\bf c}t,{\bf u}+{\bf c},t)$ with
${\bf c}=$ const is a solution of Eq.~(\ref{master}).]

For concreteness, we now consider the case where $F({\bf u};{\bf u}')$ can be
represented by the Gaussian
\begin{equation}
   F({\bf u};{\bf u}') = \left({\Lambda\over 4\pi \kappa}\right)^{3/2}
   \exp\left[-{\left({\bf u}-{\bf u}'
   +\gamma{\bf u}'/\Lambda\right)^2\over 4\kappa/\Lambda}\right]
\end{equation}
which satisfies Eq.~(\ref{constr}). In the limit $\Lambda\rightarrow\infty$,
one obtains
\begin{equation}
   \int d{\bf u}'\,F({\bf u};{\bf u}')\,g({\bf u}') - g({\bf u}) =
   \Lambda^{-1}{\cal L}_{\rm FP}g({\bf u})
\end{equation}
to leading order in $\Lambda^{-1}$. Removing the prefactor $\Lambda^{-1}$ by
means of the parameter redefinition
\begin{equation}
   \gamma/\Lambda\rightarrow\gamma\,,\quad \kappa/\Lambda\rightarrow \kappa\,,
\end{equation}
Eq.~(\ref{master}) then takes the form
\begin{eqnarray}\label{central}
   & & \left[{\partial\over\partial t}+{\bf u}\cdot\nabla_x
   +{\bf A}({\bf x})\cdot\nabla_u\right]f({\bf x},{\bf u},t) = \nonumber \\
   & & = {\cal L}_{\rm FP}\int_0^t dt'\,\Phi(t-t')\,
   {\cal P}^{t,t'}f({\bf x},{\bf u},t')\,.
\end{eqnarray}
This equation, a rigorous result based on a consideration of continuous time
random walks in phase space, is a nontrivial extension of the usual KFP equation,
Eq.~(\ref{Kramers}), which is recovered for $\Phi(\tau)=\delta(\tau)$ [corresponding to
$w(\tau)=\exp(-\tau)$]. While generalized KFP equations with memory effects have
been considered before by many authors, the retardation effect has never been included.
The latter is important, however, in order to ensure Galilean invariance, as pointed
out before. Moreover, retardation enters quite naturally in the present CTRW framework
and its physical origin is evident, given the mixed nature of the underlying physical
process -- a particle being subject to a deterministic acceleration and a series of
random kicks.

Let us now consider the important case that the memory kernel $\Phi$ in Eq.~(\ref{central})
exhibits a power law tail, i.e.
\begin{equation}
   \Phi(t-t') \approx {1\over\Gamma(\delta)}{\partial\over\partial t}
   {1\over(t-t')^{1-\delta}} \propto - {1\over(t-t')^{2-\delta}}
\end{equation}
in the long-time limit, where the prefactor is given in terms of the gamma function $\Gamma$.
A regularization procedure \cite{Balescu} then leads to the correspondence
\begin{equation}
   \int_0^t dt'\,\Phi(t-t')\,H(t') \rightarrow D_t^{1-\delta} H(t)
\end{equation}
where the so-called Riemann-Liouville operator $D_t^{1-\delta}$ is defined via
\begin{equation}
   D_t^{1-\delta} H(t) = {1\over\Gamma(\delta)}{\partial\over\partial t}
   \int_0^t {dt'\over(t-t')^{1-\delta}}\,H(t')
\end{equation}
for $0<\delta\le 1$. Consequently, Eq.~(\ref{kernel}) turns into
\begin{equation}\label{kernel2}
   {dw(t)\over dt} + D_t^{1-\delta} w(t) = 0\,.
\end{equation}
In the limit $\delta\rightarrow 1$, one gets $D_t^{1-\delta} w(t)\rightarrow w(t)$,
and Eq.~(\ref{kernel2}) has the solution $w(t)=\exp(-t)$, describing classical
relaxation behavior. However, for $\delta<1$ one obtains fractional equations instead.
A careful investigation of the regularization process actually shows that, in the
present case, it is appropriate to introduce a novel type of fractional derivative
-- which might be called ``fractional substantial derivative'' -- via
\begin{eqnarray}
   & & {\cal D}_t^{1-\delta}\,H({\bf x},{\bf u},t) = {1\over\Gamma(\delta)}
   \left[{\partial\over\partial t}+{\bf u}\cdot\nabla_x
   +{\bf A}({\bf x})\cdot\nabla_u\right]
   \nonumber \\
   & & \int_0^t {dt'\over(t-t')^{1-\delta}}\,
   e^{-(t-t')[{\bf u}\cdot\nabla_x+{\bf A}({\bf x})\cdot\nabla_u]}
   H({\bf x},{\bf u},t')\,.
\end{eqnarray}
We note that, alternatively, one may define this fractional substantial derivative
by means of its representation in Laplace space,
\begin{equation}
   {\cal D}_t^{1-\delta} \leftrightarrow [\lambda+{\bf u}\cdot \nabla_x
   +{\bf A}({\bf x})\cdot \nabla_u]^{1-\delta}\,.
\end{equation}
Using this definition, one arrives at
\begin{eqnarray}\label{Kramers2}
   & & \left[{\partial\over\partial t}+{\bf u}\cdot\nabla_x
   +{\bf A}({\bf x})\cdot\nabla_u\right] f({\bf x},{\bf u},t) = \nonumber \\
   & & = {\cal L}_{\rm FP} {\cal D}_t^{1-\delta} f({\bf x},{\bf u},t)\,,
\end{eqnarray}
a fractional generalization of the standard KFP equation, Eq.~(\ref{Kramers}).
The latter corresponds to the special case $\delta=1$. Moreover, it might be worth
pointing out that the operators ${\cal L}_{\rm FP}$ and ${\cal D}_t^{1-\delta}$
do not commute. Eq.~(\ref{Kramers2}) differs significantly
from the fractional KFP equation proposed by Barkai and Silbey \cite{BaSil}
in that it includes retardation effects through the fractional substantial
derivative ${\cal D}_t^{1-\delta}$. These nonlocal couplings in time {\em and space}
are crucial, as will be shown next.

For simplicity, let us focus on the case ${\bf A}({\bf x})=0$ in which the
position-velocity distribution with the initial condition 
$f({\bf x},{\bf u},t=0)=\delta({\bf x})\,\delta({\bf u})$
can be written as a superposition of Gaussians with varying variances:
\begin{eqnarray}\label{rep}
   f({\bf x},{\bf u},t) =
   \int d\alpha\int d\beta\int d\gamma\,w(\alpha,\beta,\gamma,t)\,[(2\pi)^3\det A]^{-1/2} & & \nonumber \\
   \times \exp\left[-\frac{q_{11}}{2}{\bf u}^2
   -q_{12} {\bf u}\cdot ({\bf x}-{\bf u}t)-\frac{q_{22}}{2}({\bf x}-{\bf u}t)^2\right]\,. \phantom{xx} & &
\end{eqnarray}
Here, $A$ denotes the matrix with the elements $A_{11}=\alpha$, $A_{22}=\gamma$, and
$A_{12}=A_{21}=\beta$; moreover, $q_{ij}$ are the elements of the inverse matrix $Q=A^{-1}$.
The function $w(\alpha,\beta,\gamma,t)$ can be viewed as a time-dependent variance distribution.
It obeys a partial differential equation with temporal memory which can be solved analytically.
For details, the reader is referred to Ref.~\cite{CTRW}.
Eq.~(\ref{rep}) can be used to calculate the long-time behavior of the second-order
moments of the distribution function $f({\bf x},{\bf u},t)$. Here, it proves useful
to distinguish two distinct situations, $\gamma=0$ and $\gamma\ne0$. For $\gamma=0$,
we obtain
\begin{eqnarray}
   \lim_{t \rightarrow \infty}\,
   \langle u(t)^2\rangle & = & \frac{2 \kappa}{\Gamma(1+\delta)}\,t^\delta\,, \\
   \lim_{t \rightarrow \infty}\,
   \langle x(t)\,u(t)\rangle & = & \frac{2 \kappa}{\Gamma(2+\delta)}\,t^{1+\delta} \,, \\
   \lim_{t \rightarrow \infty}\,
   \langle x(t)^2\rangle & = & \frac{4 \kappa}{\Gamma(3+\delta)}\,t^{2+\delta}
\end{eqnarray}
in the one-dimensional case.\cite{CTRW} The extension to multiple dimensions
is straightforward and only modifies the prefactors in the above equations.
For $\delta>0$, the system exhibits superballistic diffusion; for $\delta=1$,
it reduces to the so-called Obukhov model.\cite{Obukhov} We would like to point out that
for $\gamma=0$, the second-order moments are not affected by retardation effects,
unlike the higher-order moments. For $\gamma\ne0$, on the other hand, one gets \cite{CTRW}
\begin{eqnarray}
   \lim_{t \rightarrow \infty}\,
   \langle u(t)^2\rangle &=& \frac{\kappa}{\gamma}\,, \\
   \lim_{t \rightarrow \infty}\,
   \langle x(t)\,u(t)\rangle &=& {1\over\Gamma(2-\delta)}\,
   \frac{\kappa}{\gamma^2}\,t^{1-\delta}+(1-\delta)\,\frac{\kappa}{\gamma}\,t\,, \\
   \lim_{t \rightarrow \infty}\,
   \langle x(t)^2\rangle &=& {2\over\Gamma(3-\delta)}\,{\kappa\over\gamma^2}\,t^{2-\delta}
   +(1-\delta)\,\frac{\kappa}{\gamma}\,t^2\,.
\end{eqnarray}
For $\delta=1$, Eq.~(\ref{Kramers2}) turns into the usual KFP
equation, Eq.~(\ref{Kramers}), which is known to exhibit Gaussian statistics and
normal diffusion in the long-time limit, while for $\delta<1$, the corresponding
probability distributions are non-Gaussian. In the latter case, the terms
proprotional to $(1-\delta)$ will dominate for very long times, leading to ballistic
motion -- independent of the values of $\gamma$ and $\kappa$ as long as both are non-zero.
These terms have their origin in the collision term's inherent memory. Consequently,
retardation effects are observed to alter the diffusion properties of the system even
qualitatively.

In summary, using a CTRW-like approach, we were able to rigorously derive a new fractional
equation of the Kramers-Fokker-Planck type for particles subject to a deterministic acceleration
and a series of random kicks. In this context, we found that the generalized Fokker-Planck
collision operator necessarily involves a fractional {\em substantial} derivative,
representing important nonlocal couplings in time {\em and space}. For the force-free case,
a closed solution was given in terms of a superposition of Gaussian distributions with
varying variances. This result was then used to calculate the long-time behavior of the
second-order moments, revealing the ballistic nature of the associated diffusion processes
for any $0<\delta<1$ as long as $\gamma$ and $\kappa$ are both non-zero. This universality
could be identified as a direct consequence of retardation effects which are expressed
mathematically by means of the fractional substantial derivative. In conclusion, the
present work may be considered as a useful starting point for further investigations of
anomalous diffusion of inertial, weakly damped particles.

\end{document}